\documentclass[useAMS,usenatbib]{mn2e}

\usepackage{tabularx}
\usepackage{multirow}
\usepackage{url}
\usepackage{amsmath}
\usepackage{aas_macros}
\usepackage{graphicx}
\usepackage{subfigure}

\title[MRI: anisotropy and nonlocality]{On characterizing nonlocality and anisotropy for the magnetorotational instability}

\author[F. Nauman and E. G. Blackman]{Farrukh Nauman$^{1}$\thanks{E-mail:
fnauman@pas.rochester.edu} and Eric G. Blackman $^{1}$\thanks{E-mail: blackman@pas.rochester.edu}\\
$^{1}$Department of Physics and Astronomy, University of Rochester, Rochester, NY 14627, USA}

\begin{document}

\date{\today}

\pagerange{\pageref{firstpage}--\pageref{lastpage}} \pubyear{2014}

\maketitle

\label{firstpage}

\begin{abstract}
The extent to which angular momentum transport in accretion discs is primarily local or non-local and what determines this is an important avenue of study for understanding accretion engines. Taking a step along this path, we analyze simulations of the magnetorotational instability (MRI) by calculating energy and stress power spectra in stratified isothermal shearing box simulations in several new ways.
We divide our boxes in two regions, disc and corona where the disc is the MRI unstable region and corona is the magnetically dominated region. We calculate the fractional power in different quantities, including magnetic energy and Maxwell stresses and find that they are dominated by contributions from the lowest wave numbers. This is even more dramatic for the corona than the disc, suggesting that transport in the corona region is dominated by larger structures than the disc. By calculating averaged power spectra in one direction of $k$ space at a time, we also show that the MRI turbulence is strongly anisotropic on large scales when analyzed by this method, but isotropic on small scales. Although the shearing box itself is meant to represent a local section of an accretion disc, the fact that the stress and energy are dominated by the largest scales highlights that the locality is not captured within the box. This helps to quantify the intuitive importance
of global simulations for addressing the question of locality of transport, for which similar analyses can be performed.
\end{abstract}

\begin{keywords}
accretion, accretion discs - mhd - instabilities - turbulence.
\end{keywords}

\section{Introduction}
Angular momentum transport in accretion discs has been a long standing topic of research (for a review, see e.g. \cite{2008bhad.book.....K}, \cite{2009AnRFM..41..283S}, \cite{2011ppcd.book..283K}, \cite{2012PhyS...86e8202B}, \cite{2013LRR....16....1A}, \cite{2013arXiv1304.4879B}). Rapid variability observed in systems like Active Galactic Nuclei (AGN) hints at some kind of an enhanced mechanism for transport caused by turbulence. \cite{1973A&A....24..337S} constructed a 1-D model of an accretion disc with an $\alpha$ parametrization for such an enhanced diffusion mechanism that they used to approximate the turbulent transport. Because the formalism is of practical value, results of possible hydrodynamic and magnetohydrodynamic transport mechanisms are often quantified in terms of $\alpha$ but understanding the various mechanisms and the validity of quantifying them in such a simple parameterization are both topics of active research. One limitation of a ``local'' formalism for transport is that these accretion engines also commonly have jets and coronae which are large scale phenomena. The question of what determines the fraction of local vs. non-local transport for a general accretor is an engaging avenue of research.

The magnetorotational instability (MRI) (\cite{1991ApJ...376..214B}, \cite{1998RvMP...70....1B}) has emerged as a plausible solution to the long standing angular momentum transport problem, at least for highly ionized discs. 
The radial-azimuthal stresses from shearing box simulations, when averaged over the whole box, first show a definitive exponential growth, and subsequently saturate in a fully developed turbulent state. In interpreting the results from  simulations, it is necessary to be careful in distinguishing conceptual lessons that can be learned form specific results that may depend on the choice of initial $B$ fields, boundary conditions, domain size, and resolution. For a detailed convergence study and review of previous work of both local and global simulations see \cite{2011ApJ...738...84H}. 

While the MRI is commonly thought of as a source of local transport, an important result of stratified MRI simulations (e.g. local:  \cite{2000ApJ...534..398M}, global: \cite{2010MNRAS.408..752P}) in this context is that MRI turbulent discs lead to formation of a laminar coronal region where magnetic fields dominate thermal pressure at a few scale heights (typically 2 scale heights) above from the mid-plane thought to be the corona. The transport properties are expected to be different in the corona and the formation of such hints at the emergence of non-local processes.

The main focus of our paper is to quantitatively assess the locality and anisotropy of MRI generated turbulence by calculating energy and stress spectra for a set of stratified shearing box simulations with different domain sizes and resolution.  We are also explore whether the scale of the dominant transport structures are strongly affected by the numerical setup and the extent to which convergence among our simulations emerges. We describe our numerical setup in section 2.  In section 3, we discuss our results based on spectral calculations. We synthesize the interpretation of our results with those of previous MRI literature in section 4. We conclude in section 5.

\section{Numerical methods}
We use the publicly available finite volume high order Godunov code ATHENA (\cite{2005JCoPh.205..509G}, \cite{2008ApJS..178..137S}). We solve the ideal MHD equations in the shearing box approximation with orbital advection (\cite{2010ApJS..189..142S}). Our setup is  identical to \citep{2012MNRAS.422.2685S}: we use an isothermal equation of state $p=\rho c_s^2$, where $p = \text{pressure}, \rho = \text{density}, c_s = \text{sound speed}$. We perform density stratified simulations with a Gaussian density profile: $\rho = \rho \exp (-z^2/H^2) $ where $H = \sqrt{2} c_s/\Omega = \text{scale height}$ and $\Omega = \text{angular velocity}$. Our initial conditions are $\Omega = 10^{-3}, \rho_0 = 1, p_0 = 5 \times 10^{-7}$. We use periodic boundaries in `y' and `z' and sheared periodic boundaries in `x' \citep{1995ApJ...440..742H}.

We use a constant initial $\beta = 2 p_0/B_0^2 = 100 $ throughout the disc, with initial net toroidal field having a Gaussian profile in the vertical direction just like the density. We impose no net initial vertical field to avoid  known numerical issues  (\cite{2013ApJ...767...30B}, \cite{2013A&A...552A..71F}, \cite{2013A&A...550A..61L}).

We distinguish between disc and corona using the height above which the ratio of magnetic to thermal pressure ($\equiv \beta$) drops below unity. The magnetically dominated region with $\beta<1$ is defined as the corona while the the thermally dominated region $\beta>1$ is the disc. Computing quantities for the separate restricted regions of corona and disc means that even for a shearing box with periodic boundaries in `y' and `z' and shear periodic boundary in `x', periodicity in the vertical boundaries is lost and so we avoid the use of 3D Fourier transforms. We only Fourier transform physical quantities in `x' and `y'. We then take the complex norm squared before averaging over the relevant vertical region. 

\begin{table*}
\centering
\begin{tabular}{| l | c | c | c | c |}
\hline \hline
  & Domain & Resolution & Orbits & $\alpha$ \\ 
  & $(L_x,L_y,L_z)$ & (zones/H) & $(2\pi / \Omega)$ & $(\times 10^{-3})$ \\ \hline \hline
  24Perx2y4 & $(2,4,8)$ & 24 & 100 & $7.105 \pm 2.758$ \\
  24Perx4y2 & $(4,2,8)$ & 24 & 100 & $7.897 \pm 1.807$ \\
  24Perx4y4 & $(4,4,8)$ & 24 & 100 & $6.470 \pm 1.523$\\
  24Perx4y8 & $(4,8,8)$ & 24 & 54 & $8.899 \pm 2.062$ \\
  24Perx8y4 & $(8,4,8)$ & 24 & 100 & $5.921 \pm 0.950$ \\ \hline
  48Perx2y4 & $(2,4,8)$ & 48 & 42 & $6.347 \pm 1.549$ \\
  48Perx4y2 & $(4,2,8)$ & 48 & 51 & $5.109 \pm 0.927$ \\
  48Perx4y4 & $(4,4,8)$ & 48 & 39 & $8.259 \pm 0.929$ \\
  48Perx4y8 & $(4,8,8)$ & 48 & 43 & $8.897 \pm 2.575$ \\
  48Perx8y4 & $(8,4,8)$ & 48 & 37 & $7.445 \pm 1.113$ \\ \hline \hline
\end{tabular}
\caption{Summary of different runs. $\alpha = \frac{\langle \rho v_x v_y - B_x B_y \rangle}{p_0}$ is time averaged from 15 orbits onwards and volume averaged over the whole box.}

\end{table*}

\begin{figure}
  \centering
    \includegraphics[scale=0.5]{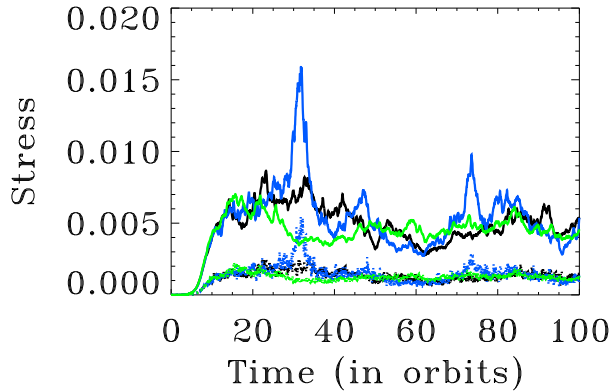}
  \caption{Time evolution of Maxwell stresses $\langle -B_x B_y \rangle/p_0$ (solid) and Reynolds stresses $\langle \rho v_x v_y \rangle/p_0$ (dotted) for 24Perx2y4 (blue), 24Perx4y4 (black), 24Perx8y4 (green). }
\label{fig:alphavst}
\end{figure}

We vary domain size and resolution for our runs to study their effects on MRI transport properties. Table 1 provides a summary of our runs and figure \ref{fig:alphavst} provides a time history of stresses. Note that we vary the domain size and the resolution separately. The shorthand label for each run is given in the first column and the last column quotes the value of Shakura-Sunyaev $\alpha = \langle \rho v_x v_y - B_x B_y \rangle/p_0$. Note that for the calculation of $\alpha$ and the turbulent spectra below, we subtract off the background shear ${\bf v} = {\bf v}_{\text{total}} + q \Omega x {\bf e}_y$, where the shear parameter $q \equiv - d \ln \Omega/d \ln r = 3/2$ for Keplerian flows.

For a quantity $F({\bf x})$ in configuration whose 2D Fourier transform is $f({\bf k})$), we compute 1D power spectra by averaging the 2D spectra in three different ways: 

(i) circle average:   $2\pi k |f(k)|^2 = \int |f({\bf k}')|^2 \delta (|{\bf k}'| - k) d{\bf k}'/\int \delta (|{\bf k}'| - k) d{\bf k}'$, where $|{\bf k}| = \sqrt{k_x^2 + k_y^2}$, 

(ii) y-average:  $|f(k_x)|^2 = \int |f({\bf k})|^2 dk_y/\int dk_y$, 

(iii) x-average:  $|f(k_y)|^2 = \int |f({\bf k})|^2 dk_x/\int dk_x$.

Note that 1D averaged energy spectra have units of energy per unit wavenumber and they are normalized by initial pressure $p_0$ (see figure \ref{fig:energy_xy}). Our stress spectra (e.g. $|B_xB_y(k)|^2$ in figure \ref{fig:stress24vs48}) plots are similar except for an important difference that stress spectra have units of stress squared per unit wavenumber and so we have normalized them by initial pressure squared. Note that the wavenumber $k \equiv 2 \pi n/L$, where $n$ is mode number, and $L$ is the length of the domain in one of the directions.

In addition to plotting power spectra, we calculate the fractional power contained in a range of wave numbers from the minimum $k_{\text{min}}$ up to $k$ to estimate of the scales of structures that dominant the energy and stresses. The fractional power $Q_f(k)$ for circle averaged power spectrum $2\pi k|f(k)|^2$ is calculated as follows:
\begin{equation}
Q_f(k) = \frac{ \int_{k_{\text{min}}}^{k} |f(k')|^2 2 \pi k' dk' } { \int_{k_{\text{min}}}^{k_{\text{max}}} |f(k')|^2 2 \pi k' dk' },
\label{fracpwrk}
\end{equation}
where $k_{\text{min}}$ and $k_{\text{max}}$ are wave numbers corresponding to the first mode and and the last mode respectively. While for `x' or `y' averaged power spectrum we use the following method:
\begin{equation}
Q_f(k_i) = \frac{ \int_{k_{i\text{min}}}^{k_i} |f(k_i')|^2 dk_i' } { \int_{k_{i\text{min}}}^{k_{i\text{max}}} |f(k_i')|^2 dk'_i }, 
\label{fracpwrkxy}
\end{equation}
where $k_i$ is either $k_x$ or $k_y$ and $k_{i\text{min}}$, $k_{i\text{max}}$ represent the first and last mode respectively. We ignore contributions from the zeroth mode, given our periodic boundaries. Note that for a given 1D averaged energy spectrum (e.g. $|B(k_x)|^2$), $Q_f(k_x)$ represents fractional power in energy as a function of radial wavenumber. However, for stress power spectra, for example $|B_xB_y(k_x)|^2$, this formula gives fractional power $Q_f(k_x)$ in units of stress squared as a function of the radial wavenumber. So for stresses, we take the fractional power to be $\sqrt{Q_f(k_x)}$.

\section{Results for MRI Energy and Stresses as a function of scale}

We calculate power spectra of energy and stresses to study the non-locality and anisotropy of MRI turbulence. Existing literature (e.g. \cite{2009ApJ...694.1010G}, \cite{2012MNRAS.422.2685S} ) has mostly focused on calculating correlation functions as a measure of non-locality. While correlations tell us about the scale over which a given quantity varies, the correlation by itself lacks information about the the magnitude of that quantity as a function of scale. Our complementary focus on the fractional power, provides key information on the strength of these structures that the MRI tends to produce.

\subsection{Length scale dependence of turbulent spectra}

\begin{figure}

  \centering
    \includegraphics[scale=0.5]{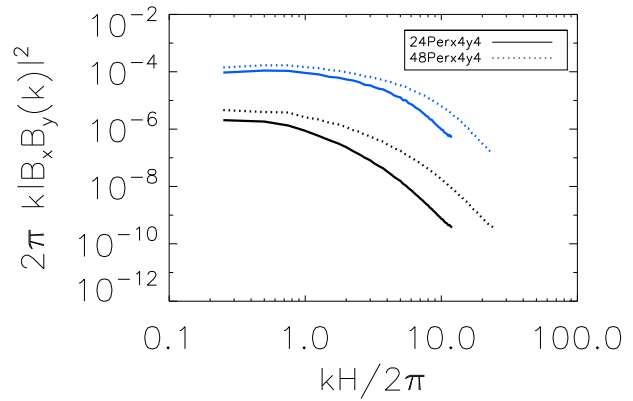}

  \caption{Resolution dependence of $xy$ Maxwell stress power spectrum multiplied by $k$ for the disc (blue) and corona (black). Note that the quantity plotted has units of stress squared per unit wavenumber. Coronal spectra for both resolutions do not show a turnover while the disc spectra do.}
  
\label{fig:stress24vs48}
\end{figure}

We plot the xy Maxwell stress spectrum in figure \ref{fig:stress24vs48} for two different resolutions. Note that since the disc and corona spectra are vertically averaged over their respective region, the total stress spectra computed by our procedure would be the weighted sum of the disc and corona and would thus lie somewhere between the two. Turnovers in the spectra are only seen in the disc region, not in the corona. Convergence is hard to define for power spectra but these plots suggest that the low wavenumber power distributions do not change with resolution. The difference is, however, pronounced at higher wavenumbers. The ratio of volume averaged xy Maxwell stress in the corona compared to the disc for the two runs 24Perx4y4 and 48Perx4y4 is 0.0316 and 0.040 respectively. We have not varied our vertical box sizes but it would ultimately interesting to compute this fraction as a function of both vertical and radial domain sizes.

\begin{figure*}

  \centering
    \includegraphics[scale=0.45]{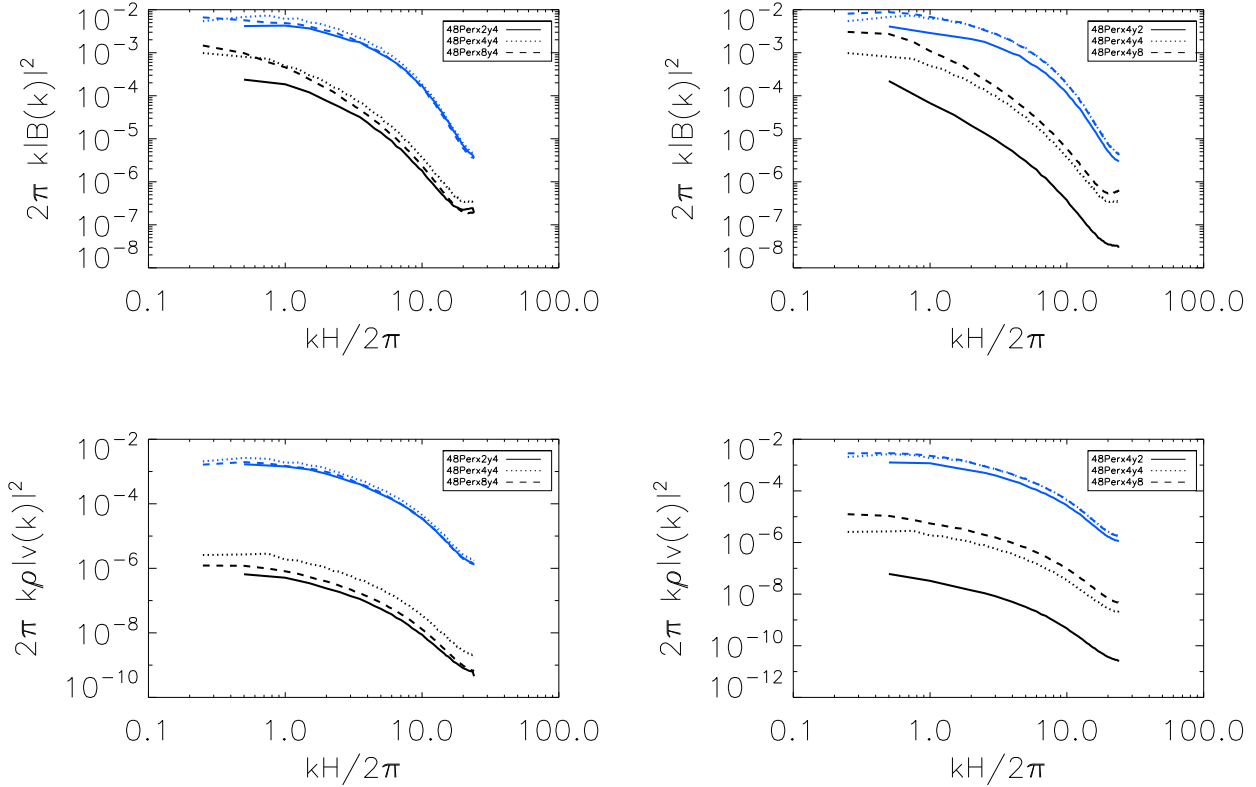}

  \caption{Domain size dependence of energy (density) spectra multiplied by $k$ for the disc (blue) and corona (black). Note that the run 48Perx4y4 represents the case of $L_x=L_y=4H$ and is plotted for cases in which we vary the radial (left) and azimuthal (right) domain sizes. For the magnetic energy spectra, the $L_x=2H,4H$ (top left) and $L_y=4H,8H$ (top right) runs show a turnover in the disc region but those for the largest radial domain $L_x=8H$ the smallest azimuthal domain $L_y=2H$ do not. Except for the smallest radial domain $L_x=2H$, both the disc and coronal kinetic energy (bottom left) spectra show a turnover. Kinetic energy spectra for $L_x=2$ and $L_y=8H$ (bottom right) do not show a turnover in the corona but both these runs have spectra in the disc region that flattens off at low wavenumbers.}
\label{fig:energy_xy}
\end{figure*}

We show plots of circle averaged kinetic energy spectra in figure \ref{fig:energy_xy}. Most of the spectral plots do not exhibit a clear peak. The exceptions are: (1) the kinetic energy power spectra for the disc and corona both show peaks for the two largest $L_x$ domains and (2) the magnetic energy spectrum shows a turnover for the disc region in the $L_x=2H$ and $L_x=4H$ run, but not for the largest run $L_x=8H$.

We observe no turnover for magnetic energy spectra in the corona for any of our runs, which indicates that while the disc (MRI dominated region) has shown some signs of the inertial range being completely captured within the domain of the simulation, the corona needs even higher resolution or even larger domain size to assess where its outer scale lies. The fact that for both kinetic and magnetic energy the few lowest wave numbers hold the most power  suggests that the shape of the spectra may not be independent of the boundary conditions. Since we use periodic boundaries, which are not physical, this raises basic questions  about convergence not only of our simulations but periodic boxes in general: can a shearing box simulation can be large enough to fully capture the essence of transport but local enough to still represent a meaningful approximation?

A noteworthy feature as we change the domain size, is the gap between disc and corona spectra. For the magnetic energy spectra, this gap narrows as we increase the domain size, particularly at lower wavenumbers while for the kinetic energy spectra the gap is considerably larger and does not change much for larger domains. The effect of this gap can also be seen in the real space volume averaged ratio of kinetic energy in the corona compared to the disc, which is for example 0.0580 for the run 48Perx4y4. For the same run, the corresponding volume averaged ratio of magnetic energy in the corona vs the disc is 0.258.

\subsection{Fractional power as function of scale and anisotropy}

\begin{figure*}

  \centering
    \includegraphics[scale=1.5]{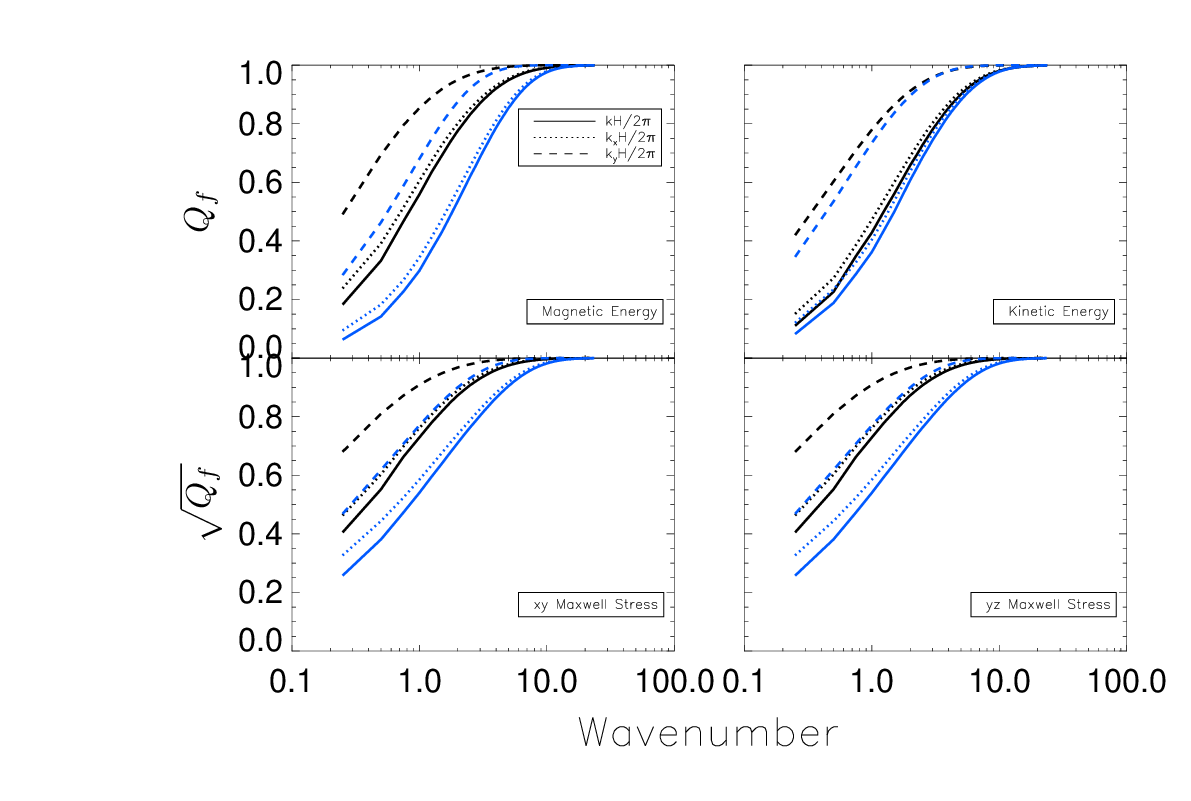}

  \caption{Fractional power for various quantities for the run 48Perx4y4 for the disc (blue) and corona (black). Note that the fractional power in both the disc and corona is the same for the kinetic energy whereas the corona has greater fractional power in the first few wavenumbers for all the magnetic quantities. The fractional power distribution in $k_y$ is significantly different from $k_x$ and $k$ for all quantities. Fractional power is calculated using equation \ref{fracpwrk} and \ref{fracpwrkxy} from the different 1D averaged spectra calculated from 2D spectrum $|f({\bf k})|^2 = |B({\bf k})|^2, |\rho v({\bf k})|^2, |B_x B_y ({\bf k})|^2, |B_y B_z ({\bf k})|^2$ for magnetic energy, kinetic energy, xy Maxwell stress, yz Maxwell stress respectively. While for the energy $Q_f$ gives the fractional power in energy, for the stresses it gives fractional power in stress squared so here we plot $\sqrt{Q_f}$.}
\label{fig:48maxwell}
\end{figure*}

The dominance of power in large scales is most dramatically conveyed by the fractional power calculations. Figure \ref{fig:48maxwell} shows fractional power of magnetic energy, kinetic energy and the two Maxwell stresses for the run 48Perx4y4. All of these quantities are dominated by the first few wavenumbers. This is something we observe across all of our runs. Calculations of such fractional power spectra do not seem to have been done in the previous literature.

In addition, we have analyzed the anisotropy by comparing 1D averaged power spectra with respect to $k_x$, $k_y$ and $k$. Like \citep{2011A&A...528A..17L}, we find that spectra exhibit strong anisotropy at the largest scales but become more isotropic at smaller scales. For all of the quantities that we have considered, the fractional power in the first few `$k_y$' modes is much greater than the corresponding fractional power in `$k_x$'. This is consistent with the notion that because the shear time scale is the same on all scales and the turnover time is shorter on small scales, the small scales are less affected by shear than the large scales. Structures which get elongated because of shear turn out to be the most dominant structures for transport. We also find that varying the domain size does not have any effect on anisotropy up to the domain sizes we have used. Given that we do not see a clear outer scale to our inertial range in any of our runs, it remains to be seen whether the anisotropy we see persists for very high resolution simulations with explicit dissipation but the physical argument above suggests that it would.

Finally, note that we also find the fractional power in kinetic energy (figure \ref{fig:48maxwell}) to exhibit the same behaviour in both the disc and corona whereas for magnetic energy, the first few lowest modes have greater fractional power relative to the total power in the corona than for the corresponding fractional power within the disc compared to the total disc power. This result is consistent with the notion that the largest magnetic structures survive MRI shredding and rise to the corona \citep{2009ApJ...704L.113B}. This also suggests that kinetic structures are rather unaffected by buoyancy. 

While we have focused on  the anisotropy of energy and stress spectra with respect to different wavenumber components, we also note that another way to characterize anisotropy is to compare power spectra of different components of the magnetic field, for example $B_x^2$, $B_y^2$ and $B_z^2$. Examples of such a  comparison for global simulations include \cite{2012ApJ...744..144F} and \cite{2014MNRAS.438.2513P}.

\section{Discussion and  Connections to Previous Work}
In the literature, we have found some examples of analysis of MRI generated turbulence using energy spectra and a few cases with stress spectra as well. In the following subsections, we compare our findings with previous work.

\subsection{Convergence and Non-Locality}

Table 1 shows that the values of $\alpha$ are similar for all of our runs. Thus if convergence of volume averaged stresses were the only important criterion, our simulations might be said to be converged. However, if one were to define a `turnover' or a unique peak in the spectra as an essential criterion for numerical convergence, or for capturing the dominant spectral range of MRI turbulence, then we would conclude that we have not achieved convergence with either resolution or domain size. In particular, although some of the disc spectra for different resolution and domain sizes do show a peak, magnetic energy spectra in the corona do not show any peak at all for any our runs. Overall however, even for cases when there a spectral peak, our energy and stress spectra calculations show that the MRI leads to predominantly non-local anisotropic transport within the shearing box.

The implication of our results is different from (but not inconsistent with) the notion that the MRI energy injection may not be local \citep{2011A&A...528A..17L}: Regardless of where the injection scale may be, the saturated MRI state seems to indicate magnetic energy migrating toward large scales. \cite{2010A&A...514L...5F} is the only paper that we have found that shows a clear low wave number turnover in magnetic energy spectra. Unlike our present simulations however, these results were for  unstratified and small radial domain boxes with explicit dissipation. Nevertheless, the fact that they see a clear turnover is significant because earlier work employing similar domain sizes but lower resolution (\cite{1995ApJ...440..742H}, \cite{1996ApJ...463..656S}) did not find such a turnover. This hints at some resolution dependence of the `non-local' structures for these unstratified boxes, even though the range of domain sizes studied was not large.

Note that both \cite{2010A&A...514L...5F} and \cite{2011A&A...528A..17L} describe unstratified MRI simulations with small box sizes. It is plausible that because of the interplay between Parker instability and MRI, the largest scales in the disc may rise to the corona \citep{2009ApJ...704L.113B}. It is also not clear whether the forcing or energy injection scale for unstratified and stratified MRI simulations should be the same. \cite{2010ApJ...713...52D} study energy spectra for different resolutions both with and without stratification and find that the spectra only converge with resolution for stratified cases.

One might ask how we can make conclusions about whether the  MRI leads to non-local transport given that shearing box is itself a local approximation. The issue is whether a shearing box can be large enough to capture the dominant  MRI  energetics and stresses while still being small enough to represent a local approximation. Indeed the distinction between local and non-local can  only be fully tested with  a global simulation that incorporates a more realistic geometry \citep{2008A&A...481...21R}. Although a thorough energy spectra analysis has not been done for global simulations, there are several examples that include energy (e.g. \cite{2012ApJ...744..144F}, \cite{2013arXiv1309.6916S}) and stress spectra (\cite{2011MNRAS.416..361B}, \cite{2012ApJ...749..189S}, \cite{2013MNRAS.435.2281P}) and they all seem to indicate that the dominant magnetic structures are large scale, and the transport non-local.

\subsection{Importance of stress spectra and torque}
 
We also point out that there has been a dearth of study of stress spectra in the literature and perhaps even more fundamentally, the torque.  Since the problem of disc transport revolves around stresses, understanding the dominant scales of angular momentum transport are bolstered by MRI stress spectra in addition to energy. For example, \cite{2013MNRAS.435.2281P} show that while magnetic energy spectra `converge' for all of their runs, their radial stress spectrum for the lowest resolution run does not. Other than our results herein (figure \ref{fig:48maxwell}, bottom right), we are unaware of an previous studies of the fractional power of the vertical stress spectrum and we find that its power distribution is nearly identical to the radial stresses (figure \ref{fig:48maxwell}, bottom left). But this needs to be explored with higher resolution simulations. 

In shearing box simulations, vertical stresses are typically an order of magnitude or more smaller (\cite{2000ApJ...534..398M}) when averaged compared to radial stresses. But a vertical stress smaller than a radial stress by a ratio of $H/R$ could still produce a torque comparable to that from the radial stress. It is indeed torque that matters for angular momentum transport. Since vertical stresses are likely associated with coronae and jets their importance for assessing the non-locality of MRI induced transport is further exacerbated. 

The nature of outflows generated from a MRI unstable shearing box configuration in the presence of a net vertical field has been the subject of recent work by \cite{2009ApJ...691L..49S}, \cite{2013A&A...552A..71F}, \cite{2013ApJ...767...30B}, \cite{2013A&A...550A..61L}. Due to the periodic boundary conditions used, the initial net vertical flux remains conserved throughout  the simulation. In particular, \cite{2013ApJ...767...30B} find that the ratio of radial and vertical transport depends sensitively on the initial field strength.  The shearing box setup includes radial shear but no vertical shear, a limitation in the interpretation of the azimuthal-vertical stresses as compared to real discs. Nonetheless \cite{2013ApJ...767...30B} seem to suggest indeed that vertical transport can become comparable with radial transport for $\beta \sim 1000$ if $H/R \sim 0.1$.

\section{Conclusion}
Using energy and stress power spectra and fractional power as a function of wave number, we have demonstrated that the MRI leads to  predominantly anisotropic and non-local turbulent structures at least within a shearing box. These findings are broadly consistent with previous MRI studies that employ correlation functions. However, our computation of the fractional power spectrum in energy and stress   helps assess the question of locality  more directly than previous approaches.  We find that not only that MRI leads to non-local structures but that these structures dominate transport. Because our simulations were conducted for a shearing box, we cannot assess how nonlocal the  transport would be in a global simulation from the MRI but the method we have used to assess this would be applicable. Our findings based on study of a range of domain sizes do suggest that the anisotropy and non-locality would likely persist on non-local scales within a global disc. An important physical implication of this result is to highlight that the transport found in MRI simulations may not be congruent with the local, isotropic and radial-only transport model of \cite{1973A&A....24..337S}, further motivating the opportunities to improve the basic semi-analytic framework for modeling accretion discs.

\bibliography{general}

\begin{thebibliography}{}

\bibitem[\protect\citeauthoryear{{Abramowicz} \& {Fragile}}{{Abramowicz} \&
  {Fragile}}{2013}]{2013LRR....16....1A}
{Abramowicz} M.~A.,  {Fragile} P.~C.,  2013, Living Reviews in Relativity, 16,
  1

\bibitem[\protect\citeauthoryear{{Bai} \& {Stone}}{{Bai} \&
  {Stone}}{2013}]{2013ApJ...767...30B}
{Bai} X.-N.,  {Stone} J.~M.,  2013, \apj, 767, 30

\bibitem[\protect\citeauthoryear{{Balbus} \& {Hawley}}{{Balbus} \&
  {Hawley}}{1991}]{1991ApJ...376..214B}
{Balbus} S.~A.,  {Hawley} J.~F.,  1991, \apj, 376, 214

\bibitem[\protect\citeauthoryear{{Balbus} \& {Hawley}}{{Balbus} \&
  {Hawley}}{1998}]{1998RvMP...70....1B}
{Balbus} S.~A.,  {Hawley} J.~F.,  1998, Reviews of Modern Physics, 70, 1

\bibitem[\protect\citeauthoryear{{Beckwith}, {Armitage} \& {Simon}}{{Beckwith}
  et~al.}{2011}]{2011MNRAS.416..361B}
{Beckwith} K.,  {Armitage} P.~J.,    {Simon} J.~B.,  2011, \mnras, 416, 361

\bibitem[\protect\citeauthoryear{{Blackman}}{{Blackman}}{2012}]{2012PhyS...86e8202B}
{Blackman} E.~G.,  2012, \physscr, 86, 058202

\bibitem[\protect\citeauthoryear{{Blackman} \& {Pessah}}{{Blackman} \&
  {Pessah}}{2009}]{2009ApJ...704L.113B}
{Blackman} E.~G.,  {Pessah} M.~E.,  2009, \apjl, 704, L113

\bibitem[\protect\citeauthoryear{{Blaes}}{{Blaes}}{2013}]{2013arXiv1304.4879B}
{Blaes} O.,  2013, ArXiv e-prints

\bibitem[\protect\citeauthoryear{{Davis}, {Stone} \& {Pessah}}{{Davis}
  et~al.}{2010}]{2010ApJ...713...52D}
{Davis} S.~W.,  {Stone} J.~M.,    {Pessah} M.~E.,  2010, \apj, 713, 52

\bibitem[\protect\citeauthoryear{{Flock}, {Dzyurkevich}, {Klahr}, {Turner} \&
  {Henning}}{{Flock} et~al.}{2012}]{2012ApJ...744..144F}
{Flock} M.,  {Dzyurkevich} N.,  {Klahr} H.,  {Turner} N.,    {Henning} T.,
  2012, \apj, 744, 144

\bibitem[\protect\citeauthoryear{{Fromang}}{{Fromang}}{2010}]{2010A&A...514L...5F}
{Fromang} S.,  2010, \aap, 514, L5

\bibitem[\protect\citeauthoryear{{Fromang}, {Latter}, {Lesur} \&
  {Ogilvie}}{{Fromang} et~al.}{2013}]{2013A&A...552A..71F}
{Fromang} S.,  {Latter} H.,  {Lesur} G.,    {Ogilvie} G.~I.,  2013, \aap, 552,
  A71

\bibitem[\protect\citeauthoryear{{Gardiner} \& {Stone}}{{Gardiner} \&
  {Stone}}{2005}]{2005JCoPh.205..509G}
{Gardiner} T.~A.,  {Stone} J.~M.,  2005, Journal of Computational Physics, 205,
  509

\bibitem[\protect\citeauthoryear{{Guan}, {Gammie}, {Simon} \& {Johnson}}{{Guan}
  et~al.}{2009}]{2009ApJ...694.1010G}
{Guan} X.,  {Gammie} C.~F.,  {Simon} J.~B.,    {Johnson} B.~M.,  2009, \apj,
  694, 1010

\bibitem[\protect\citeauthoryear{{Hawley}, {Gammie} \& {Balbus}}{{Hawley}
  et~al.}{1995}]{1995ApJ...440..742H}
{Hawley} J.~F.,  {Gammie} C.~F.,    {Balbus} S.~A.,  1995, \apj, 440, 742

\bibitem[\protect\citeauthoryear{{Hawley}, {Guan} \& {Krolik}}{{Hawley}
  et~al.}{2011}]{2011ApJ...738...84H}
{Hawley} J.~F.,  {Guan} X.,    {Krolik} J.~H.,  2011, \apj, 738, 84

\bibitem[\protect\citeauthoryear{{Kato}, {Fukue} \& {Mineshige}}{{Kato}
  et~al.}{2008}]{2008bhad.book.....K}
{Kato} S.,  {Fukue} J.,    {Mineshige} S.,  2008, {Black-Hole Accretion Disks
  --- Towards a New Paradigm ---}

\bibitem[\protect\citeauthoryear{{K{\"o}nigl} \& {Salmeron}}{{K{\"o}nigl} \&
  {Salmeron}}{2011}]{2011ppcd.book..283K}
{K{\"o}nigl} A.,  {Salmeron} R.,  2011, {The Effects of Large-Scale Magnetic
  Fields on Disk Formation and Evolution}.
pp 283--352

\bibitem[\protect\citeauthoryear{{Lesur}, {Ferreira} \& {Ogilvie}}{{Lesur}
  et~al.}{2013}]{2013A&A...550A..61L}
{Lesur} G.,  {Ferreira} J.,    {Ogilvie} G.~I.,  2013, \aap, 550, A61

\bibitem[\protect\citeauthoryear{{Lesur} \& {Longaretti}}{{Lesur} \&
  {Longaretti}}{2011}]{2011A&A...528A..17L}
{Lesur} G.,  {Longaretti} P.-Y.,  2011, \aap, 528, A17

\bibitem[\protect\citeauthoryear{{Miller} \& {Stone}}{{Miller} \&
  {Stone}}{2000}]{2000ApJ...534..398M}
{Miller} K.~A.,  {Stone} J.~M.,  2000, \apj, 534, 398

\bibitem[\protect\citeauthoryear{{Parkin}}{{Parkin}}{2014}]{2014MNRAS.438.2513P}
{Parkin} E.~R.,  2014, \mnras, 438, 2513

\bibitem[\protect\citeauthoryear{{Parkin} \& {Bicknell}}{{Parkin} \&
  {Bicknell}}{2013}]{2013MNRAS.435.2281P}
{Parkin} E.~R.,  {Bicknell} G.~V.,  2013, \mnras, 435, 2281

\bibitem[\protect\citeauthoryear{{Penna}, {McKinney}, {Narayan},
  {Tchekhovskoy}, {Shafee} \& {McClintock}}{{Penna}
  et~al.}{2010}]{2010MNRAS.408..752P}
{Penna} R.~F.,  {McKinney} J.~C.,  {Narayan} R.,  {Tchekhovskoy} A.,  {Shafee}
  R.,    {McClintock} J.~E.,  2010, \mnras, 408, 752

\bibitem[\protect\citeauthoryear{{Regev} \& {Umurhan}}{{Regev} \&
  {Umurhan}}{2008}]{2008A&A...481...21R}
{Regev} O.,  {Umurhan} O.~M.,  2008, \aap, 481, 21

\bibitem[\protect\citeauthoryear{{Shakura} \& {Sunyaev}}{{Shakura} \&
  {Sunyaev}}{1973}]{1973A&A....24..337S}
{Shakura} N.~I.,  {Sunyaev} R.~A.,  1973, \aap, 24, 337

\bibitem[\protect\citeauthoryear{{Shariff}}{{Shariff}}{2009}]{2009AnRFM..41..283S}
{Shariff} K.,  2009, Annual Review of Fluid Mechanics, 41, 283

\bibitem[\protect\citeauthoryear{{Simon}, {Beckwith} \& {Armitage}}{{Simon}
  et~al.}{2012}]{2012MNRAS.422.2685S}
{Simon} J.~B.,  {Beckwith} K.,    {Armitage} P.~J.,  2012, \mnras, 422, 2685

\bibitem[\protect\citeauthoryear{{Sorathia}, {Reynolds}, {Stone} \&
  {Beckwith}}{{Sorathia} et~al.}{2012}]{2012ApJ...749..189S}
{Sorathia} K.~A.,  {Reynolds} C.~S.,  {Stone} J.~M.,    {Beckwith} K.,  2012,
  \apj, 749, 189

\bibitem[\protect\citeauthoryear{{Stone} \& {Gardiner}}{{Stone} \&
  {Gardiner}}{2010}]{2010ApJS..189..142S}
{Stone} J.~M.,  {Gardiner} T.~A.,  2010, \apjs, 189, 142

\bibitem[\protect\citeauthoryear{{Stone}, {Gardiner}, {Teuben}, {Hawley} \&
  {Simon}}{{Stone} et~al.}{2008}]{2008ApJS..178..137S}
{Stone} J.~M.,  {Gardiner} T.~A.,  {Teuben} P.,  {Hawley} J.~F.,    {Simon}
  J.~B.,  2008, \apjs, 178, 137

\bibitem[\protect\citeauthoryear{{Stone}, {Hawley}, {Gammie} \&
  {Balbus}}{{Stone} et~al.}{1996}]{1996ApJ...463..656S}
{Stone} J.~M.,  {Hawley} J.~F.,  {Gammie} C.~F.,    {Balbus} S.~A.,  1996,
  \apj, 463, 656

\bibitem[\protect\citeauthoryear{{Suzuki} \& {Inutsuka}}{{Suzuki} \&
  {Inutsuka}}{2009}]{2009ApJ...691L..49S}
{Suzuki} T.~K.,  {Inutsuka} S.-i.,  2009, \apjl, 691, L49

\bibitem[\protect\citeauthoryear{{Suzuki} \& {Inutsuka}}{{Suzuki} \&
  {Inutsuka}}{2013}]{2013arXiv1309.6916S}
{Suzuki} T.~K.,  {Inutsuka} S.-i.,  2013, ArXiv e-prints

\end{thebibliography}
\bibliographystyle{mn2e}

\section*{Acknowledgments}

We thank R. Parkin for pertinent  comments and discussion. FN acknowledges Horton Fellowship from the Laboratory for Laser Energetics at U. Rochester and we acknowledge support from NSF grant AST-1109285. We thank Karim Shariff for several discussions about locality of turbulence that helped clarify important issues. We are grateful to Shane Davis for graciously providing his Fourier transform routines that served as the starting point for our data analysis. We acknowledge the Center for Integrated Research Computing at the University of Rochester for providing computational resources.

\bsp

\label{lastpage}

\end{document}